# Chiral symmetry on the lattice

Michael Creutz[a] [*]

[a]Physics Department, Brookhaven National Laboratory, PO Box 5000, Upton, NY 11973-5000, USA
creutz@wind.phy.bnl.gov

I review some of the difficulties associated with chiral symmetry in the context of a lattice regulator. I discuss the structure of Wilson Fermions when the hopping parameter is in the vicinity of its critical value. Here one flavor contrasts sharply with the case of more, where a residual chiral symmetry survives anomalies. I briefly discuss the surface mode approach, the use of mirror Fermions to cancel anomalies, and finally speculate on the problems with lattice versions of the standard model.

## 1. INTRODUCTION

Chiral symmetry has long been an important concept in particle physics. The issues range from the partially conserved axial current in strong interaction physics to the parity violation in the electroweak sector of the standard model.

On the other hand, the subject has been particularly frustrating when discussed in the context of a lattice regulator. At this meeting there are six parallel sessions directly on these issues, in addition to numerous relevant talks scattered through the other sessions. This is much more material than I can possibly understand, let alone review, so in this talk I will limit myself to a few old issues.

The bulk of my discussion concerns what happens at the critical value of the hopping parameter for Wilson Fermions. I first discuss the one flavor case, where the anomaly breaks the chiral symmetry completely. I then contrast this with the case of two flavors, where a residual flavor-non-singlet axial symmetry survives when the Fermion mass goes to zero.

Having done one and two flavors, I leave the $N$ flavor case as a homework problem. This is actually a cheat, since the doublers in the Wilson approach come in multiplets, and the structure with multiple flavors is in fact relevant to even the one flavor case.

I will make heavy use of the analogy between electrodynamics in two space-time dimensions and the non-Abelian gauge theories of quarks and gluons in four dimensions. These theories both exhibit confinement, chiral symmetry breaking, and the emergence of a corresponding $CP$ violating parameter. This analogy was emphasized some time ago by Kogut and Susskind [1], but the ideas are well worth revisiting.

One useful aspect of this analogy is that the two dimensional electrodynamics case is amenable to study via bosonization [5]. This allows a simple semiclassical analysis of the behavior of the theory as a function of the Fermion mass.

One new feature I use is the role of surface modes to the structure of the theory. While these play a key role in the Kaplan [17] approach to chiral lattice theories, the concept is useful in understanding the conventional Wilson theory as well. Indeed, such states appear naturally as the hopping parameter passes through its critical value.

At the end I make some general comments on the surface mode approach to a chiral theory, and some of the problems associated with using this for a formulation of the standard model on the lattice.

## 2. CHIRAL SYMMETRY

The concepts of chiral symmetry are deeply entwined with the Lorentz group. Indeed, the representation structure of this group is qualitatively different for massless versus massive particles. Only in the massless case is the helicity of a particle invariant under boosts and rotations, and is separating helicity components a Lorentz invariant concept. When a spin 1/2 Fermion is coupled minimally to a gauge theory, this helicity conservation survives interactions, and currents associated with left and right handed particles are naively separately conserved. The Fermion fields naturally break into two independent parts, $\psi_R$ and $\psi_L$.

With only one spatial dimension the concept is even simpler. A massless excitation travelling at the speed of light can never be overtaken. Thus a particle moving to the right does so in all frames, and the fields break up into left and right moving parts.

The infamous chiral anomalies complicate this simple picture. Indeed, with gauge interactions present, not all axial currents can be conserved. This lies at the root of much fascinating physics, entwined with such issues as strong $CP$ violation, the presence of an unanticipated parameter $\theta$, the so called $U(1)$ problem, the mass of the $\eta$ meson, etc.

The anomaly arises from a sliding of states in and out of the Dirac sea[4]. Indeed, the necessity of anomalies follows intuitively from simple band theory. Consider a massive Fermion with energy spectrum $E = \sqrt{p^2 + m^2}$. In the Dirac picture, there is also a filled sea of negative energy states, in which holes represent antiparticles. The Fermi level of the vacuum lies at $E = 0$, between these two bands, and we have the classic picture of an insulator. On the other hand, if we consider a massless Fermion with $E = \pm p$, there is no gap in the spectrum. In this case we expect the vacuum to be a conductor. Now when electromagnetic fields are applied to a conductor, currents are induced. In one space dimension, the electric current is the difference of the number of right moving and left moving particles, which therefore cannot be conserved. Indeed, without anomalies, transformers would not work.

## 3. MASS TERMS

An effect of the anomaly arises when we consider the usual Fermion mass term in the Lagrangian density

$$\mathcal{L}_m = m\overline{\psi}\psi \tag{1}$$

If we now consider a change of variables

$$\psi \longrightarrow e^{i\theta\gamma_5/2}\psi \tag{2}$$

the mass term becomes

$$\mathcal{L}_m \longrightarrow M\ \overline{\psi}\psi + M_5\ i\overline{\psi}\gamma_5\psi \tag{3}$$

where

$$\begin{aligned} M &= m\cos(\theta) \\ M_5 &= m\sin(\theta) \end{aligned} \tag{4}$$

Under this rotation the kinetic and gauge coupling terms of the continuum Lagrangian density are naively invariant.

As this is just a change of variables, one might conclude that physics would be equivalent if we start with either of the two forms of $L_m$ above. However, because of the chiral anomaly, with the latter form a dependence on the parameter $\theta$ survives. This is a naively unexpected new $CP$ violating parameter on which the physics of the model can depend. In what follows I explore the expected phase structure in the $(M, M_5)$ plane.

How can this $\theta$ dependence survive, when all I have done is a simple change of variables? The answer has to do with the regulation of the ultraviolet divergences to obtain a finite renormalized theory. With a Paul-Villars regulator, $\theta$ represents a relative $\gamma_5$ phase between the fundamental Fermion and the heavy regulator fields. As the regulator mass goes to infinity, a residual dependence on the angle $\theta$ survives. One of the points of this talk is that the doublers in the lattice theory play this same role of determining the phase under chiral rotations.

## 4. ONE FLAVOR IN THE CONTINUUM

To see how this dependence comes about in a specific model, let me appeal to the continuum massive Schwinger model with one flavor,

i.e. massive electrodynamics in one spatial dimension. Coleman [5] discussed this model in some detail in the context of bosonization. In one space dimension there is a precise operator correspondence between Fermion and bosonic theories. Some of the relevant mappings are

$$:\overline{\psi}\psi: \longleftrightarrow -C:\cos(2\sqrt{\pi}\phi):$$
$$:\overline{\psi}\gamma_5\psi: \longleftrightarrow -C:\sin(2\sqrt{\pi}\phi): \qquad (5)$$
$$j_\mu =:\overline{\psi}\gamma_\mu\psi: \longleftrightarrow \frac{1}{\sqrt{\pi}}\epsilon_{\mu\nu}\partial_\nu\phi$$

Here the constant $C$ depends on the scheme used to normal order the operators to remove ultraviolet divergences. Colons denote this ordering. From now on I will ignore these technical details and just use this mapping to obtain a qualitative feeling for the structure of the theory.

Note that one of these equations represents Gauss's law

$$\partial_1\phi = \sqrt{\pi}j_0; \qquad (6)$$

so, we can identify $\phi$ as proportional to the electric field (a scalar in one space dimension).

This mapping indicates that two dimensional electrodynamics is equivalent to a scalar field theory with the potential

$$V(\phi) = \frac{e^2}{2\pi}\phi^2 - mC\cos(2\sqrt{\pi}(\phi - \theta)) \qquad (7)$$

The first piece is the electromagnetic energy and the second comes from the fermion mass term. Here I have kept the $\theta$ dependence from the generalized Fermion mass term of Eq. (3). This parameter determines the relative phase between the oscillations and the parabolic part of the potential.

The potential in Eq. (7) consists of uniform oscillations added to a parabolic shape. The critical observation is that if I pick $\theta = \pi$ and if $m$ is large enough, then this potential will have a classic double well form, as sketched in Figure 1. This is the sign of a first order phase transition, and we conclude that the model has a phase diagram in the $(M, M_5)$ plane as qualitatively sketched in Figure 2.

Along the transition line following the negative $M$ axis, the field $\phi$ has a non-vanishing expectation value. This suggests that $\sin(\phi)$ also does, and thus we are led to the expectation value $\langle\overline{\psi}\gamma_5\psi\rangle$ as a natural order parameter in the Fermionic representation. This quantity should be discontinuous across the phase transition line. As this is a pseudoscalar, the transition represents a spontaneous breaking of parity symmetry. Ref. [5] presents a beautiful physical interpretation of the parity breaking phase in terms confinement and so called "half-asymptotic states."

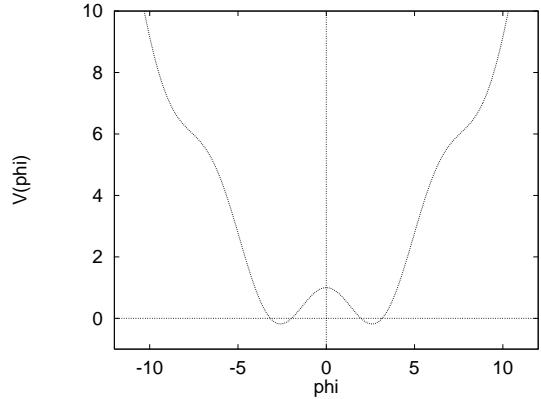

Figure 1. The effective potential for the massive Schwinger model at $\theta = \pi$. The units are arbitrary.

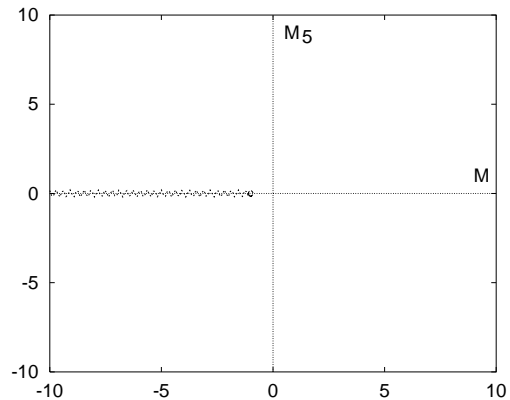

Figure 2. The phase diagram for the one flavor case, in arbitrary units. The wiggly line represents a first order phase transition.

Note that the transition does not occur if the magnitude of $m$ is too small. The oscillations from the mass term must overcome the quadratic part of the potential. Indeed, with one flavor the point $m = 0$ is not special (except that the Schwinger model happens to be exactly solvable there). The anomaly has succeeded in breaking all aspects of chiral symmetry.

Note that if we redefine the scalar field $\phi$ by adding a constant, one can move the $\theta$ dependence between the quadratic "gauge" term and the "Fermion" mass term

$$\begin{aligned}V(\phi) = & \frac{e^2}{2\pi}\phi^2 - mC\cos(2\sqrt{\pi}(\phi-\theta)) \\ \longleftrightarrow & \frac{e^2}{2\pi}(\phi+\theta)^2 - mC\cos(2\sqrt{\pi}\phi)\end{aligned} \quad (8)$$

In four dimensions we expect a similar possibility of switching between a term in the Lagrangian density of form $\theta F\tilde{F}$ and generalizing the mass term to include a piece of form $M_5\overline{\psi}\gamma_5\psi$. Indeed, given the difficulties with defining the former form on the lattice [6], it may be simpler instead to use the latter.

For later purposes, it is instructive to consider the model in a finite box with open boundaries. If I consider the case where the field $\phi$ has a background expectation value, but use boundary conditions that $\phi$ vanishes on the system walls, then there will be regions around the boundaries where the field adjusts itself to settle into a minimum of the potential. From Gauss's law, Eq. (6), in this region there will be a buildup of non-vanishing charge. Indeed, as sketched in Figure 3, the system automatically creates a surface charge which serves to generate the background field in the interior of the system.

Changing the boundary condition to some fixed non-zero value for $\phi$ can be thought of as apply-ing an external field to the system, which in turn modifies the amount of surface charge generated. The above shifting of $\theta$ dependence between the "gauge" and "Fermion" parts of the action represents changing the relative balance between the surface charge and an external applied field. I raise these issues of surface charge because surface modes also arise quite naturally in Wilson's lattice Fermion approach, to which I now turn.

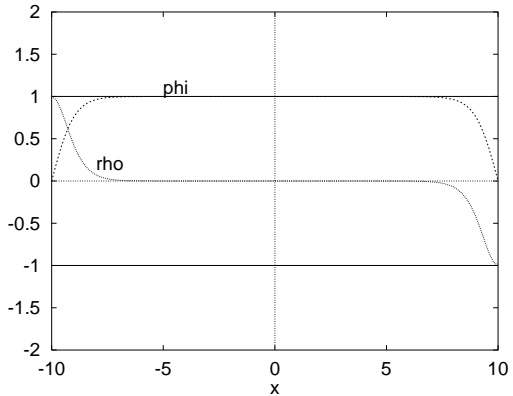

Figure 3. With open boundaries, charge appears on the ends of the system and generate the background field. The charge density is the derivative of the field settling to its bulk value.

## 5. WILSON FERMIONS

I treat the lattice Fermions in a Hamiltonian approach, primarily to assist with quantum mechanical intuition. In one space dimension, consider the Hamiltonian

$$\begin{aligned}H(K,r,M) = \sum_j & K\left(\overline{\psi}_j(i\gamma_1+r)\psi_{j+1}\right) \\ & + K\left(\overline{\psi}_{j+1}(-i\gamma_1+r)\psi_j\right) \\ & + M\overline{\psi}_j\psi_j\end{aligned} \quad (9)$$

The Fermionic operator $\psi$ is a two component spinor satisfying the lattice anticommutation relations

$$[\psi_i^\dagger,\psi_j]_+ = \delta_{ij}. \quad (10)$$

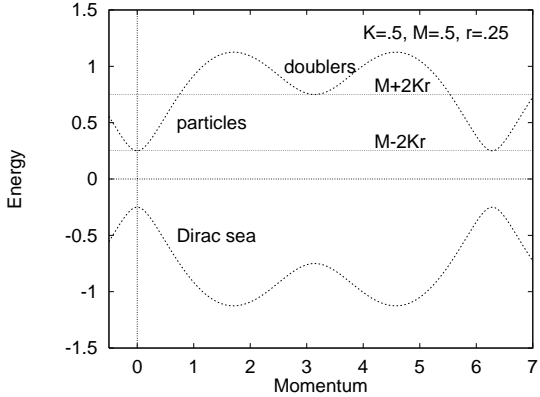

Figure 4. The spectrum of Wilson Fermions.

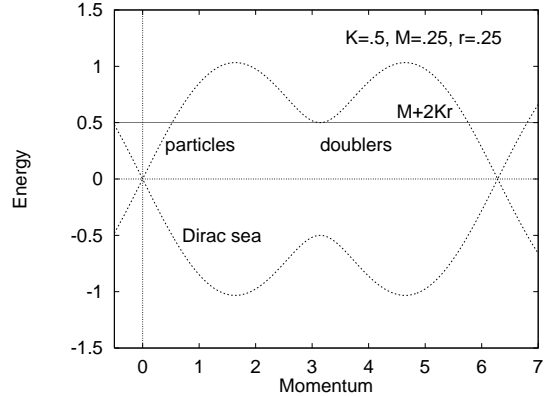

Figure 5. The spectrum of Wilson Fermions with critical hopping parameter.

The quantities $M$ and $K$ I call the mass and hopping parameters, respectively. The third parameter, $r$, represents the term Wilson [7] added to the naive discretization of the Dirac Hamiltonian to remove doublers.

As a reminder of how this works, the single particle spectrum for $H$, found by Fourier transform, is

$$E^2 = 4K^2 \sin^2(q) + (M - 2Kr\cos(q))^2 \qquad (11)$$

where $q$ is the spatial momentum. The positive energy solutions are the particles, while the negative energy states are to be filled as a Dirac sea. This is sketched in Figure 4. Note how the Wilson term separates the doubler states at momentum near $\pi$ from the low energy particle states at small momentum. The mass of the particles is $m = M - 2Kr$ and the mass of the doublers is $m_d = M + 2Kr$.

For completeness I note that if r is large enough, then the point in the spectrum near momentum $\pi$ is not actually a minimum. This is the case even with the popular choice $r = 1$, where a projection operator appears whenever a Fermion hops from one site to a neighbor. This, however, is not important for the following discussion.

To study a massless Fermion, one can tune the parameters to $M = 2Kr$. The spectrum in this case is shown in Figure 5. I use the word "tune" here because when gauge fields are included, the parameters are renormalized, and the critical value of the hopping is an apriori unknown function of the gauge coupling. This is how simulations with Wilson Fermions proceed: several values of the hopping parameter are studied, and the critical hopping found via extrapolation to the point where the pion mass vanishes.

Note that this is not the only point where the theory is critical. The mass gap in the spectrum also vanishes at $M = -2Kr$, where the doublers become massless. In three space dimensions the situation is somewhat more complex, with critical points at $M \in \{\pm 6Kr, \pm 2Kr\}$. For the case $M = 2Kr$ massless particles then occur whenever any one of the three components of momentum is near $\pi$. This gives effectively three degenerate flavors, a situation I relegated to homework at the beginning of this talk.

## 6. SURFACE MODES

Returning to the one space dimension case, something rather interesting happens already in the free Fermion situation when $M$ lies between the two critical points $\pm 2Kr$ where the Fermions or the doublers are massless. Alternatively stated, whenever $|M| < |2Kr|$, the hopping parameter is supercritical. As the hopping increases to such a value, the gap in the single particle spectrum closes and reopens. In 1939 Shockley [8] analyzed a similar situation, and argued that, with open boundaries, as the gap reopens the two lowest energy levels separate off as surface states. In Figure 6, taken from ref. [9], I sketch the spectrum of states on a 20 site lattice as a function of the hopping parameter in a case where the critical hopping is unity. Note the two levels which separate out. The energies of these levels are split because of mixing between the lattice ends; for a semi-infinite lattice at supercritical $K$ there is a surface mode at exactly zero energy.

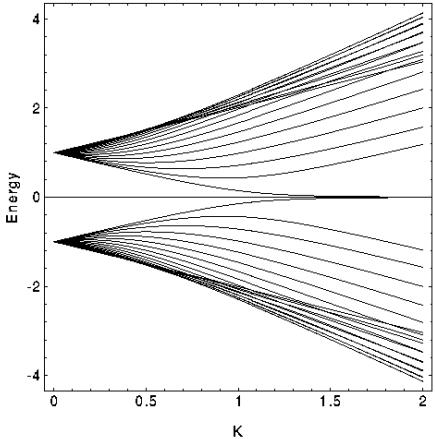
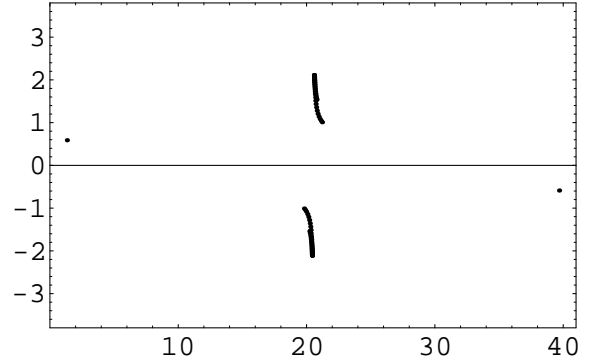

Figure 6. The spectrum of Wilson Fermions on a 20 site lattice as a function of the hopping parameter. Here I take $M = 1$ and $r = 0/5$. The critical hopping occurs at $K = 1$.

Figure 7. The energy spectrum of Wilson Fermions on a 40 site lattice as a function of the expectation of the spatial coordinate. Note the modes near the lattice boundaries.

As in the continuum, it is convenient to introduce a term $M_5 i\overline{\psi}\gamma_5\psi$ in our Hamiltonian. In this case the plane wave spectrum is modified to

$$E^2 = M_5^2 + 4K^2 \sin^2(q) + (M - 2Kr\cos(q))^2 \quad (12)$$

The analysis in the appendix of Ref. [9] is easily generalized to show that for any value of $M_5$ surface modes are present whenever $|M| < |2Kr|$. For an isolated surface the mode acquires energy $E = \pm M_5$. In Figure 7 I plot the energy spectrum versus the expectation value of the spatial coordinate for the single particle states of this Hamiltonian on a 40 site lattice with a small value of $M_5$ and $M$ in the critical region. Note the two surface modes near the ends of the lattice. The $M_5$ term breaks parity symmetry; so, the surface level on one end is lifted relative to the other.

Suppose we now fill the negative energy states. In this case there is one extra particle on one end of the system. When the gauge fields are turned on, this will generate a background field that will flow towards the empty surface state on the opposite wall. This leads us to expect the same parity violating phase as discussed earlier in terms of surface charges in the continuum picture. I am immediately led to the phase diagram for the $(M, M_5)$ plane sketched in Figure 8.

Thus a parity violating phase transition is expected in the small $M$ region with the order parameter being $\langle \overline{\psi}\gamma_5\psi \rangle$. Such a phase was briefly mentioned in ref. [10] and was extensively discussed in ref. [11]. The physics of this phase corresponds directly to $\theta = \pi$.

Several comments are in order. First, for three-space-dimensional non-Abelian models there will be considerably more structure in the interior of this diagram. While I expect an analogous critical endpoint near $M = 6Kr$, three flavors of doublers will appear at $M = \pm 2Kr$, as discussed above. This, of course, was part of your homework problem.

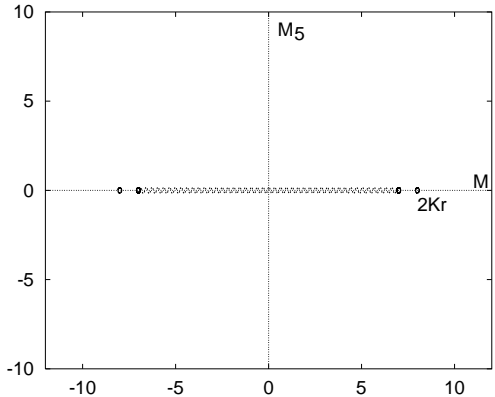

Figure 8. The phase diagram for one flavor of Wilson Fermions. The wiggly line represents a first order phase transition.

Second, while I have shown the figure in terms of $M$, the connection with the hopping parameter $K$ is reciprocal. Subcritical $M$ corresponds to supercritical $K$.

Third, some time ago Seiler and Stamatescu [12] discussed the parameter $\theta$ in terms of a $\gamma_5$ phase in the Wilson term $r$ rather than $M$. Their discussion is equivalent to that here, with $\theta$ being a relative $\gamma_5$ phase between the $M$ and $r$ terms.

Fourth, it is perhaps amusing that in this one flavor case one can make the mass of the pseudoscalar meson vanish by going to the critical endpoint of the parity violating phase transition. With only one flavor the only pseudoscalar is an analogue of the $\eta$ meson. To make the $\eta$ massless requires playing the positive contribution of the anomaly against a negative bare mass for the Fermion constituents. This is not expected to be possible with more than one flavor.

Finally, I note that a recent paper has studied hopping parameter expansions for the behavior of this critical point in the massive lattice Schwinger model as a function of the gauge coupling. [13].

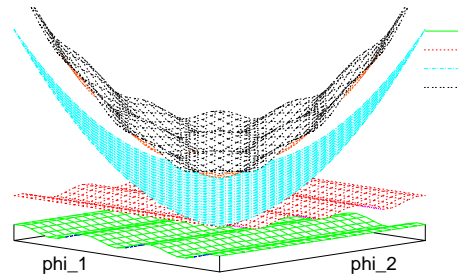

Figure 9. The potential for the two flavor case. The $(\phi_1, \phi_2)$ origin is in the center of the valley.

## 7. TWO FLAVORS IN THE CONTINUUM

Returning to the continuum case, bosonization of two Fermionic flavors requires two scalar fields [5]. The mass terms for the two are independent, but the gauge field term receives contributions from both species. Thus we are led to an effective potential

$$V(\phi_1, \phi_2) = \frac{e^2}{2\pi}(\phi_1 + \phi_2)^2 \\ - mC \cos\left(2\sqrt{\pi}(\phi_1 - \theta_1)\right) \\ - mC \cos\left(2\sqrt{\pi}(\phi_2 - \theta_2)\right) \quad (13)$$

I have included two angles for separate chiral rotations of the two fields. The first term represents a parabolic valley, while the other two are oscillations along lines at an angle of $\pm\pi/4$ with respect to the bottom of the valley. The three components and their sum are generically plotted in Figure 9.

The low lying meson spectrum can be qualitatively understood from this picture. The $\eta$ meson represents oscillations in the $\phi_1 + \phi_2$ direction, while the $\pi_0$ represent oscillations in the orthogonal $\phi_1 - \phi_2$ direction. The fact that the $\eta$ is heavier from the pion is due to the extra term proportional to $e^2$ going up the sides of the valley. The charged pions are solitons representing states interpolating between two adjacent minima in the bottom of the valley. This particular bosonization hides the underlying flavor symmetry which ensures the degeneracy of the charged and neutral pions.

The angles $\theta_{1,2}$ represent shifts of the parabolic valley relative to the oscillations from the mass terms. This valley is, however, invariant under shifts along its bottom, i.e. shifts which hold $\phi_1 + \phi_2$ constant. This corresponds to the non-anomalous flavored chiral symmetry under rotations of the Fermionic field by $\exp(i\theta\gamma_5\tau_3)$, where $\tau_3$ is the third isospin matrix.

Unlike in the single flavor case, a vanishing Fermion mass is now indeed a special case. The residual chiral symmetry allows one to rotate $m$ to $-m$, and thus physics should be symmetrical about the $M_5$ axis.

If I consider surface charges as discussed earlier, with two flavors each field independently generates its own modes. The total electric field will be the sum of the contributions from each species. In particular, the physical background field is $\theta = \theta_1 + \theta_2$.

For the phase diagram in the $(M, M_5)$ plane, if I give the same values to these parameters for each field, then the physical field will be twice as strong as in the one flavor case. Then we expect the first order phase transition representing vacuum breakdown at $\theta = \pi$ will run up the $M_5$ axis.

To clarify this picture, it is convenient to add a small amount of flavor breaking to the theory. If this breaking is put into both the $M$ and $M_5$ terms, the singularity at the origin breaks apart into two transitions from the independent flavors, as sketched in Figure 10. Very near one of the endpoints the physics looks much like the one flavor case, but as one moves along the first order line, the second flavor strengthens the field and bends the transition toward the $M_5$ axis. When the flavor breaking is removed, the chiral limit is pinched between two second order endpoints.

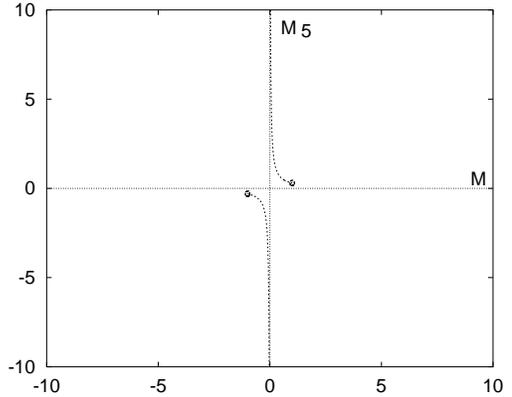

Figure 10. The phase diagram for the two flavor continuum case with a small flavor breakings in both $M$ and $M_5$. The curves follow first order phase transitions. Note how the chiral limit is pinched between two second order endpoints.

## 8. TWO FLAVORS ON THE LATTICE

It is now natural to extend this picture to the lattice with its doublers. With two flavors, each will contribute to the background field as $M_5$ is turned on. In addition, the doublers will contribute with the opposite sign, further bending the $\theta = \pm\pi$ phase transition lines. For weak couplings this combining of fields should be approximately linear

$$\theta \sim \theta_1 + \theta_2 - \theta_{d1} - \theta_{d2} \qquad (14)$$

where the various angles are those in the $(M, M_5)$ geometrically determined from the points where the various particles or doublers go massless. The resulting phase diagram is sketched in Figure 11. Note that this picture requires the model not be quenched; the fields of the various species enhance each other.

The situation is more complicated for three space dimensions due to the twelve extra doublers in the interior of the diagram of Figure 11. Nevertheless, I still conclude that the physics is qualitatively equivalent just above and below the chiral point. Unlike in the continuum case, the critical value for $M$ will be renormalized because the massive doublers break the non-anomalous axial symmetries.

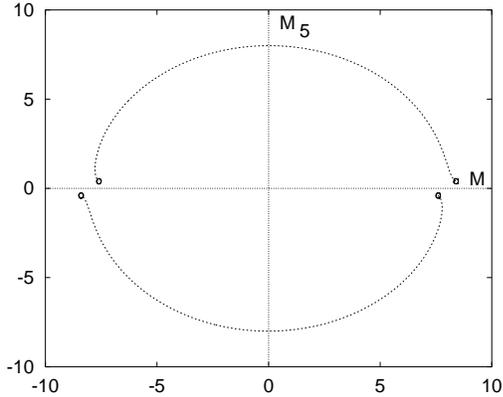

Figure 11. The conjectured phase diagram for the two flavor case on the lattice. I include a small flavor breaking in both $M$ and $M_5$ to split apart the transition at the critical endpoints. The curves represent first order phase transitions.

Things might become still more complicated for strong gauge couplings. In Ref. [11] arguments were made for a phase with spontaneous breaking of both parity and flavor. A similar phase has also been investigated for strongly coupled Nambu-Jona-Lasinio fields with Wilson Fermions [14]. It may be that such a phase arises out of the interior of the diagram in Figure 11 as the coupling is increased.

Extending this picture to four space-time dimensions, one might ask whether the nontrivial dependence on the parameter $\theta$ will survive the continuum infinite-volume limit. Indeed, for the case of electrodynamics it is generally believed that it does not (independent of the fact that QED might not even have a continuum limit). The disappearance of $\theta$ is usually ascribed to a combination of the absence of instantons and the inability of the vacuum to sustain a background field against pair production.

For four dimensional non-Abelian theories it is generally believed that a non-vanishing value for $\theta$ is possible, with instanton phenomena generating nonperturbative effects. Nevertheless, some time ago Tudron and I [15] speculated that confinement might be inconsistent with the observability of $\theta$. We suggested that the surface fields arising from a non-vanishing $\theta$ would be confined into an unobservably small region of space. This idea that confinement might automatically solve the strong $CP$ problem has resurfaced with the recent suggestion of Schierholz [16] that a continuum limit with non-vanishing $\theta$ might force one into a Higgs type phase rather than one with confinement.

## 9. THE SURFACE MODE APPROACH

Recently there has been considerable activity in using Shockley surface states as the basis for a theory of chiral Fermions [17]. For an extensive review see Ref. [18]. The idea is to set up a theory in one extra dimension so that surface modes exist, and our observed world is an interface with our quarks and leptons being these surface modes. Particle hole symmetry naturally gives the basic Fermions zero mass. We would be unaware of the extra dimension, which requires large energy to penetrate. In this picture, opposing surfaces carry states of opposite helicity, and the anomalies are due to a tunnelling through the extra dimension.

Ref. [9] discussed the general conditions for surface modes to exist. Normalized solutions are bound to any interface separating a region with supercritical from subcritical hopping. Kaplan's original paper [17], considered $M = M_{cr} + m\epsilon(x)$, where $M_{cr}$ is the critical value for the mass parameter. Shamir [19] presented a somewhat simpler picture where the hopping vanishes on one side, which then drops out of the problem.

To couple gauge fields to this theory without adding lots of unneeded degrees of freedom, it is simplest to only have gauge fields in the physical directions. In this approach, the extra dimension is perhaps best thought of as a flavor space [20]. With a finite lattice this procedure gives equal couplings of the gauge field to the Fermion modes on opposing walls in the extra dimension. The result is an effective light Dirac Fermion. In the case of the strong interactions, this provides an elegant scheme for a natural chiral symmetry without the tuning inherent in the usual Wilson approach. The breaking of chiral symmetry arises only through finiteness of the extra dimension.

The main problem at this point is that the two helicities on opposite walls are both present, so we do not have a truly chiral gauge theory as needed to describe the weak interactions. The unresolved question is whether there is some scheme to make the opposite helicity states on one wall either decouple or move to large mass.

## 10. MIRROR FERMIONS

Actually, if we use a Higgs mechanism to generate masses for our Fermions, it is quite simple to obtain a parity violating gauge theory in this surface mode picture. The idea is merely to make the Higgs coupling different on the two walls, so that the different helicity states have different masses. Such a scheme has been discussed in slightly different contexts in Refs. [9,21].

If we use this approach for a lattice theory of the weak interactions, it is not quite the usual standard model, since there exist heavy right handed partners of the light left handed neutrinos. Indeed, it is a particular realization of a mirror Fermion model, similar in many respects to a variation on Wilson Fermions proposed by Montvay [23].

One thing that appears to be impossible is to take the mirrors to infinite mass before taking a continuum limit. Ref. [22] shows that if the Higgs coupling generating the mirror particle mass is taken to infinity on the offending wall, then a multitude of doubler states descends to become yet further unwanted zero modes.

One obvious question is how heavy can the extra mirror states be made. In general, one might expect bounds on their masses, similar to the famous bounds on the Higgs particle mass. Especially if we consider a theory where the chiral anomalies are not cancelled among the light particles, the mirror particles are required for perturbative consistency. When the anomalies do cancel, as in the standard model, then there is no perturbative need for the mirrors, and the bounds on the mirror masses might be expected to be weaker. Nevertheless, such a model still has an exact baryon symmetry, with instantons taking baryons into their mirrors [24]. Thus this model cannot display the baryon non-conservation phenomenon discussed by 't Hooft [25].

## 11. THE STANDARD MODEL

Placing the standard model on the lattice remains an unsolved problem. The difficulties are apparently tied with the subtle way that anomalies cancel between the quarks and the leptons, as manifested in the famous relation between their electric charges

$$3Q_u + 3Q_d + Q_e + Q_\nu = 0. \qquad (15)$$

Accepting 't Hooft's argument that instanton phenomena violate baryon conservation, any successful lattice scheme must include this interplay between the quarks and the leptons. Any attempt to set up an electroweak theory for the leptons alone is doomed to fail.

In the surface mode approach, anomalies correspond to enhanced tunnelling through the extra dimension. For such a formulation of the standard model to work, the quarks must turn into leptons in the process. This strongly suggests that we should be working in some unified scheme. Frolov and Slavnov, [26] have been exploring closely related mechanisms based on the group $SO(10)$. Some time ago Eichten and Preskill [27] suggested explicitly adding terms to the theory which violated the known anomalous symmetries. Perhaps such terms can be used to generate masses for the mirror particles in the surface mode scheme.

Several attempts to put the standard model on the lattice involve a breaking of gauge invariance in the initial formulation, with the hope that the appropriate symmetries appear in the continuum limit [28]. Variations of this are possible in the surface mode approach as well [9,21]. However, given the elegance of the original Wilson [29] lattice gauge theory, I feel it would be a shame if we must be driven to such an extreme. Fermion violation can, however, appear quite naturally in such schemes, appearing as states slide out of the Dirac sea [30]. For relatively smooth gauge fields the resulting gauge non-invariance appears to in-volve only the high energy part of the spectrum, and optimistically may become irrelevant in the continuum limit.

## 12. CONCLUDING REMARKS

Chiral symmetry has provided a conundrum since the beginning of the lattice approach. Recent advances, to a large extent stimulated by the surface mode approach, have clarified somewhat the situation with vector theories, such as in the strong interactions.

Puzzles still remain for the weak interactions, and one might wonder if the lattice is trying to tell us something deep. Perhaps the standard model as it stands is not complete. It may be brash to elude our problems with speculations, but perhaps mirror Fermions must exist. If they have masses enough larger than the top quark, the effects on radiative corrections would be small enough not to have been observed. We should certainly be looking for them and trying to understand better the bounds on their masses.

## 13. ACKNOWLEDGEMENTS

I am quite grateful to Ivan Horvath and Shailesh Chandrasekharan for extensive discussions on the ideas presented here. I also am indebted to Ivan for the spectra appearing in the figures.

## REFERENCES

1. J. Kogut and L. Susskind, Phys. Rev. D11 (1975) 3594.
2. S. Coleman, Annals Phys. 101 (1976) 239.
3. D. Kaplan, Phys. Lett. B288 (1992) 342.
4. B. Holstein, Am. J. Phys. 61 (1993) 142; J. Ambjorn, J. Greensite, and C. Peterson, Nucl. Phys. B193 (1983) 381.
5. S. Coleman, Ann. Phys. (NY) 101 (1976) 239.
6. M. Chu, J. Grandy, S. Huang, and J. Negele, Phys. Rev. D49 (1994) 6039.
7. K. Wilson, in *New Phenomena in Subnuclear Physics*, Editied by A. Zichichi (Plenum Press, NY, 1977).
8. W. Shockley, Phys. Rev. 56 (1939) 317.
9. M. Creutz and I. Horvath, Phys. Rev. D50 (1994) 2297.
10. J. Smit, Nucl. Phys. B175 (1980) 307.
11. S. Aoki, Nucl. Phys. B314 (1989) 79; S. Aoki and A. Gocksch, Phys. Rev. D45 (1992) 3845.
12. E. Seiler and I. Stamatescu, Phys. Rev. D25 (1982) 2177.
13. H. Gausterer and C. Lang, hep-lat-9408018.
14. S. Aoki, S. Boetcher, and A. Gocksch, Phys. Lett. B331 (1994) 157; K. Bitar and P. Vranas, Phys. Rev. D50 (1994) 3406; Nucl. Phys. B, Proc. Suppl. 34 (1994) 661.
15. M. Creutz and T. Tudron, Phys. Rev. D16 (1977) 2978.
16. G.Schierholz, hep-lat-9403012 (1994)
17. D. Kaplan, Phys. Lett. B288 (1992) 342; M. Golterman, K. Jansen, D. Kaplan, Phys. Lett. B301 (1993) 219.
18. K. Jansen, hep-lat-9410018.
19. Y. Shamir, Nucl. Phys. B406 (1993) 90.
20. R. Narayanan and H. Neuberger, Phys. Lett. B302 (1993) 62; Phys. Rev. Lett. 71 (1993) 3251; Nucl. Phys. B412 (1994) 574.
21. M. Golterman, K. Jansen, D. Petcher, and J. Vink, Phys. Rev. D49 (1994) 1606.
22. M. Golterman and Y. Shamir, hep-lat-9409013.
23. I. Montvay, Nucl. Phys. B (Proc. Suppl.) 30 (1993) 621; Phys. Lett. 199B (1987) 89.
24. J. Distler and S. Rey, Princeton hep-lat-9305026.
25. G. 't Hooft, Phys. Rev. Lett. 37, 8 (1976); Phys. Rev. D14 (1976) 3432.
26. S. Frolov and A. Slavnov, Nucl. Phys. B411 (1994) 647.
27. E. Eichten and J. Preskill, Nucl. Phys. B268 (1986) 179.
28. A. Borrelli, L. Maiani, G. Rossi, R. Sisto and M. Testa, Nucl. Phys. B333 (1990) 335; J. Alonso, Ph. Boucaud, J. Cortés, and E. Rivas, Phys. Rev. D44 (1991) 3258.
29. K. Wilson, Phys. Rev. D10 (1974) 2445.
30. W. Bock, J. Hetrick, and J. Smit, hep-lat-9406015.